\documentclass[prl,twocolumn,showpacs,superscriptaddress,nofootinbib]{revtex4}
\usepackage{graphicx,color,amsmath,amssymb}
\usepackage[]{hyperref}

\def\ket#1{| #1 \rangle}
\def\bra#1{\langle #1 |}

\def\ketbra#1#2{| #1 \rangle\!\langle #2 |}
\def\II{1\!\mathrm{l}}

\begin{document}

\title{Characterization of complex quantum dynamics \\ with a scalable NMR information processor}

\author{C.A. Ryan }
\affiliation{Institute for Quantum Computing, University of Waterloo, Waterloo, ON, N2L 3G1, Canada.}

\author{J. Emerson}
\affiliation{Institute for Quantum Computing, University of Waterloo, Waterloo, ON, N2L 3G1, Canada.}
\affiliation{Perimeter Institute for Theoretical Physics, Waterloo, ON, N2J 2W9, Canada}

\author{D. Poulin}
\affiliation{Institute for Quantum Computing,  University of Waterloo, Waterloo, ON, N2L 3G1, Canada.}
\affiliation{Perimeter Institute for Theoretical Physics, Waterloo, ON, N2J 2W9, Canada}
\affiliation{School of Physical Sciences, The University of Queensland, QLD4072, Australia}

\author{C. Negrevergne}
\affiliation{Institute for Quantum Computing,  University of Waterloo, Waterloo, ON, N2L 3G1, Canada.}

\author{R. Laflamme}
\affiliation{Institute for Quantum Computing, University of Waterloo, Waterloo, ON, N2L 3G1, Canada.}
\affiliation{Perimeter Institute for Theoretical Physics, Waterloo, ON, N2J 2W9, Canada}

\date{\today}

\begin{abstract}
We present experimental results on the measurement of fidelity decay under contrasting system dynamics using a nuclear magnetic resonance quantum information processor.   The measurements were performed by implementing a scalable circuit in the model of deterministic quantum computation with only one quantum bit.   The results show measurable differences between regular and complex behaviour and for complex dynamics are faithful to the expected theoretical decay rate.  Moreover, we illustrate how the experimental method can be seen as an efficient way for either extracting coarse-grained information about the dynamics of a large system, or measuring the  decoherence rate from engineered environments.
\end{abstract}


\maketitle

Quantum information processors (QIP) promise efficient solutions to problems that seem intractable using classical devices.  Tremendous progress has been achieved in understanding how much more powerful quantum mechanics is at manipulating information  \cite{Shor:2004a}.  Some first steps have been taken in experimentally demonstrating both algorithms \cite{Vandersypen:2001a} and simulating physical systems \cite{Negrevergne:2005a}, in small prototype devices, and extensive research is pushing forward many avenues in the search for a fully scalable quantum computer \cite{QIP:2004a}.    On the way to larger and scalable devices, benchmarking and characterizing will play a crucial role in understanding the performance of devices and pointing out directions where improvement is required.   An obvious characterizing procedure is full state or process tomography \cite{Wienstein:2004a}; however, these methods will become unfeasible as we move to larger QIP.  The number of experiments necessary for tomography grows exponentially with the size of the system and the experimental time required will quickly become prohibitive.  We must turn instead, to characterizations which provide coarse-grained information.  These experiments will extract the information necessary for quantum control or error correction processes, yet will be achievable using a reasonable amount of resources - i.e. in a scalable manner.  

One such measure of interest is the fidelity decay.  This tool is useful in two contexts: characterizing the complexity of a system's dynamics; and, in studying the effects of the environment dynamics on decoherence rates.  Simulating quantum dynamics, whether of the system of interest or of a large environment, on a classical computer is inherently difficult; so, a quantum computer is a natural tool in the study of complex (chaotic) quantum dynamics \cite{Geogeot:2001a}.  Previous experimental work on the simulation of quantum chaos \cite{Weinstein:2002a} on a QIP relied on full state tomography, 	which is an inefficient, and hence unscalable means of measuring signatures of the system's complexity.
Here, we experimentally investigate complex dynamics through a  fidelity decay measurement that requires only one pseudo-pure bit of quantum information, is fully scalable in the absence of noise, and whose running time is limited only by the inherent complexity of the map we are characterizing. 

In classical mechanics, the distinction between regular and chaotic behaviour is clear:  the rate at which two states, initially close in phase space, are driven apart,  is either exponential or sub-exponential.  This measure has no relevance for isolated quantum systems, since their unitary evolution preserves the distance between two states.  The field of quantum chaos looks at other signatures of chaos \cite{Haake:2001a}.   Fidelity decay under perturbation, initially proposed by Peres \cite{Peres:1984a},  is an important analogue to the classical notion of instability, and has received much study \cite{Jalabert:2001a,Jacquod:2001a,Benenti:2002a,Emerson:2002a}.  Rather than considering the overlap of two nearby states evolving under the same evolution, this method considers the overlap of the same initial state evolved under two nearby unitary maps.  If the discrete time evolution is given by the map $U$, then we also consider the map $U_p = UP$ where $P = \exp\left( -iV\right)$ for some hermitian matrix $V$.   Then, the fidelity decay after $n$ steps is given by,
\begin{equation}
\label{FDeqn}
F_n\left(\psi\right) = \left|\bra{\psi}\left(U^n\right)^\dagger U_p^n\ket{\psi}\right|^2.
\end{equation}

The analogy with classical mechanics suggests that chaotic systems might show an exponential fidelity decay, while regular systems would decay at a slower (e.g. polynomial) rate.  However, numerical studies of small systems show a much more complicated situation,  with different behaviours 	depending on the form and strength of the perturbation \cite{Jacquod:2001a,Benenti:2002a,Emerson:2002a}. Yet, for a range of sufficiently strong perturbations, chaotic systems show a universal exponential fidelity decay at an average rate which is given by the Fermi Golden Rule (FGR) \cite{Emerson:2002a,Emerson:2005b}.  The average decay is 	universal, in the sense that it is independent of the system dynamics and depends only on certain details of the perturbation.  Specifically, the rate of the universal exponential decay depends only on the variance of the eigenvalues of the perturbation (i.e., the 2-norm of the perturbation, $\| V \|_2$) and continues until a saturation level $\mathcal{O}(\frac{1}{N})$, due to the finite system dimension $N$. Hence, direct measurement of the fidelity decay under different applied perturbations can provide important information about the complexity of the system dynamics. 

\begin{figure}[htbp]
\includegraphics[scale=0.47]{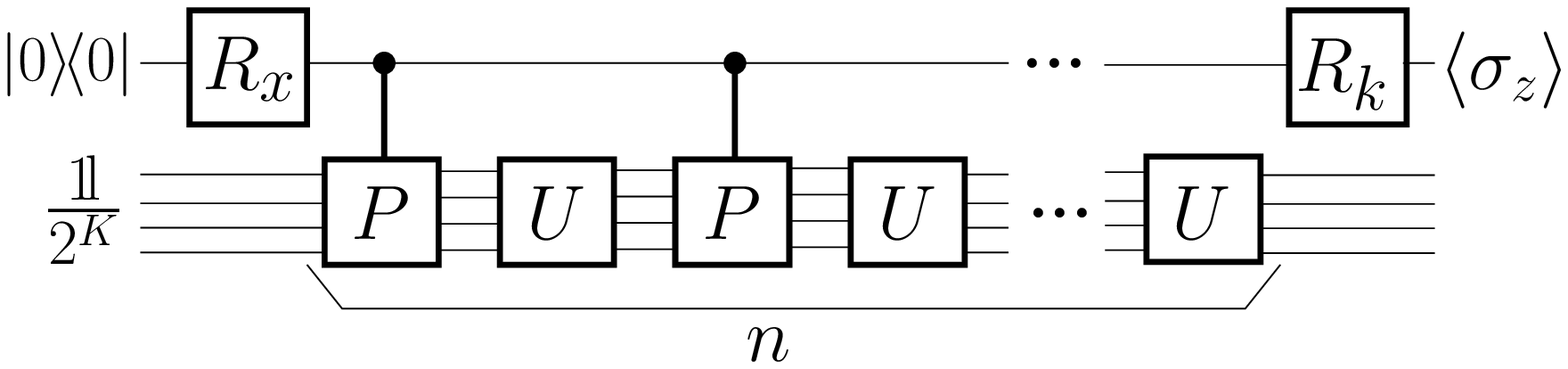}
\caption{\label{ideal_circuit}
Ideal quantum circuit for measuring fidelity decay after n steps.  The top qubit in a pure state can be considered either as a probe of the bottom system of $K$ qubits which starts in the maximally mixed state or as a toy system being decohered by the maximally mixed environment below.  Making the final rotation about the $x$ ($y$ ) basis gives the real (imaginary) part of the trace in Eq. \ref{traceeqn}.  Taken from Ref. \cite{Poulin:2004a}} 
\end{figure}

Given some dynamical system under study, the fidelity decay measured for any one particular initial state will fluctuate from the average - an effect 	particularly pronounced with small systems.  Hence estimating the average decay can require averaging over many different initial states.  However, recently \citeauthor{Poulin:2004a} \cite{Poulin:2004a} worked out a quantum circuit (see Fig. \ref{ideal_circuit})  for directly measuring the average fidelity decay within the deterministic quantum computation with a single bit (DQC1) model \cite{Knill:1998a}.  The fidelity decay after $n$ steps, averaged over a uniform (Haar) measure of the initial states $\ket{\psi}$ (see Eq. \ref{FDeqn}) for an arbitrary system of dimension $N$ takes the general form,
\begin{equation}
\label{traceeqn}
\overline{F(n)} = \frac{\left\vert Tr\left\{\left(U^n\right)^\dagger\left(PU\right)^n\right\}\right\vert^2 + N}{N^2 + N}.
\end{equation}

Therefore, experimental determination of the average fidelity decay for a particular $U$ and $P$ requires measuring the trace, for which there exists an efficient DQC1 circuit \cite{Miquel:2002a}. 

Liquid state NMR offers a test-bed technology with a sufficient number of qubits and control for the first demonstration of a DQC1 algorithm.  The DQC1 model, although more limited than full quantum computation,  is interesting, since it questions entanglement as the basic concept behind the extra-power of quantum computers \cite{Caves:2005a}.   Although the model does not allow for error correction, the required initial state is not a full multi-qubit pseudo-pure state as in conventional NMR QIP \cite{Vandersypen:2004a}, but a \emph{single} pseudo-pure qubit of the form  $\rho_\epsilon = \frac{1-\epsilon}2 \II + \epsilon\ketbra 00$, with all other qubits starting in the maximally mixed state. This state is very close to the thermal state of liquid state NMR and can be efficiently prepared from the thermal distribution. 

\begin{figure}[htbp]
\includegraphics[scale = 0.7]{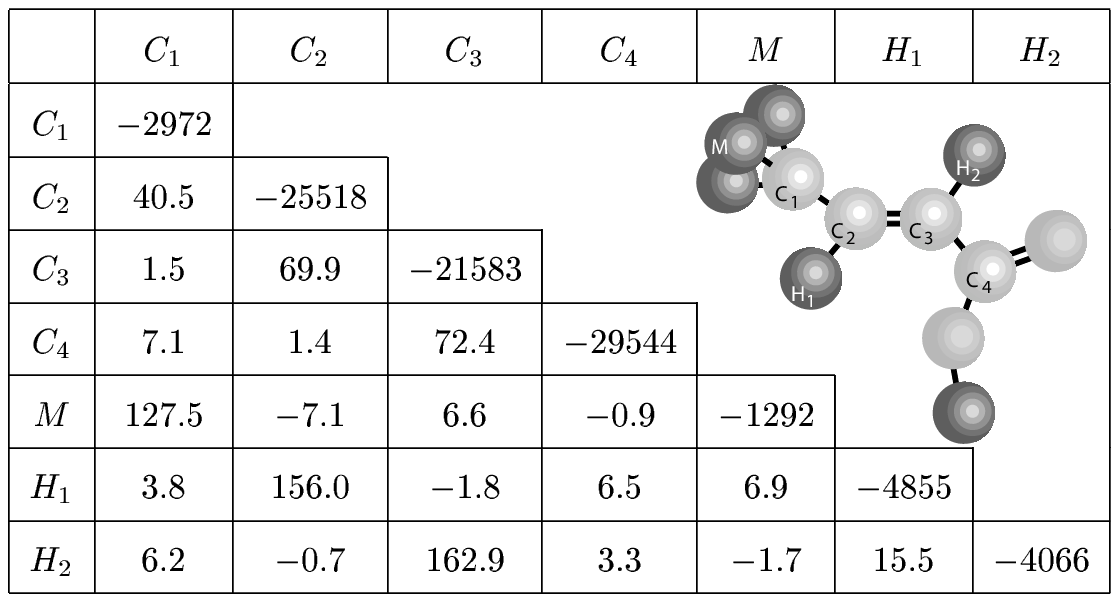}
\caption{\label{ca}
The molecule trans-crotonic acid and its natural Hamiltonian - diagonal terms are chemical shifts and off-diagonal elements coupling terms in Hz.  The darkly shaded nuclei are hydrogen and the lightly, carbon.  The unlabeled nuclei are oxygen whose natural abundance have zero spin.  Hence, we can ignore their coupling with the rest of the molecule.  The three hydrogen nuclei of the methyl group are magnetically equivalent.  They form a composite spin, from which we can select the spin $\frac{1}{2}$ subspace, giving us an additional qubit.  }
\end{figure}

The experiments were implemented on a Bruker Avance 700 MHz spectrometer using the molecule trans-crotonic acid (shown in Fig. \ref{ca}).  The four carbons were used as the system qubits and the spin $\frac{1}{2}$ selected part of the methyl group was the probe.  This setup allowed the use of broadband refocussing pulses to decouple the probe qubit from the system qubits during the application of the random map $U$.  $H_1$ and $H_2$ were decoupled by placing them in a pseudo-pure state at the beginning of the pulse sequence.  Measurement was performed by quadrature detection of the free induction decay of the methyl group which gives both $\left<\sigma_x\right>$ and $\left<\sigma_y\right>$.  Further details of the molecule and experimental technique can be found in Ref. \cite{Knill:2000a}.    

We investigated the difference in fidelity decay response for regular evolution - provided by the natural Hamiltonian of the molecule - versus complex evolution - achieved by using our ability to generate arbitrary dynamics (universal control) to implement a pseudo-random map \cite{Haake:2001a}.  We used a decomposition for pseudo-random operators  based on repeated applications of a two step process: (1) Individual random rotations are applied to each qubit; (2) Simultaneous two-body interactions between neighbours given by the unitary, $U = \exp \left[ i\left(\frac{\pi}{4}\right)\sum_{j=1}^{n-1}\sigma^{j}_{z}\otimes\sigma^{j+1}_{z}\right]$ \cite{Emerson:2003a}.  For a finite number of repetitions of these two steps, the resulting distribution of maps is biased with respect to the uniform measure; however, it converges exponentially to the uniform measure \cite{Emerson:2005a}.   For this experiment the first step was replaced by correlated random rotations, induced by broadband pulses of random phase and power on either side of a randomly chosen delay period providing random rotations about the z-axis and some two body interactions.  In order to reach a suitable trade-off between decoherence effects and the randomness of the map four iterations were implemented.  Although there is not as many random parameters in our experimental scheme as in the theoretical procedure, numerical simulations showed that, after four repetitions it implemented a sufficiently random map to simulate complex dynamics.  The first group to implement these pseudo-random operators measured their degree of randomness through complete state tomography \cite{Emerson:2003a}. Note that here, from the measurement of the fidelity decay we can efficiently characterize the degree of randomness of the map.

The controlled perturbation was implemented using the natural Hamiltonian of the molecule that provides a coupling between the probe and the system qubits of the form: $\sum_{j=1}^4J_{MC_i}\sigma_{z}^{M}\otimes\sigma^{Ci}_{z}$. This can been seen as a controlled rotation of the system qubits about the z axis.  Therefore, a controlled operation on the system qubits can be implemented by allowing the natural evolution for a time proportional to the desired strength of the perturbation.  This perturbation can be transformed into a rotation about another axis, by sandwiching the evolution with two rotations of the target qubits.  The fidelity decay depends on the eigenvector statistics of the system in the eigenbasis of the perturbation \cite{Emerson:2002a}.  For the large majority of randomly chosen perturbations, a regular system will have random statistics in that eigenbasis and show an exponential decay at the FGR rate.  However, for perturbations given by simple functions of a system coordinate, only the integrable system will have a structured form in the perturbation eigenbasis and hence exhibit a fidelity decay with large fluctuations from the FGR average.

\begin{figure}[ht!]
$\begin{array}{c}
\includegraphics[scale = 0.45]{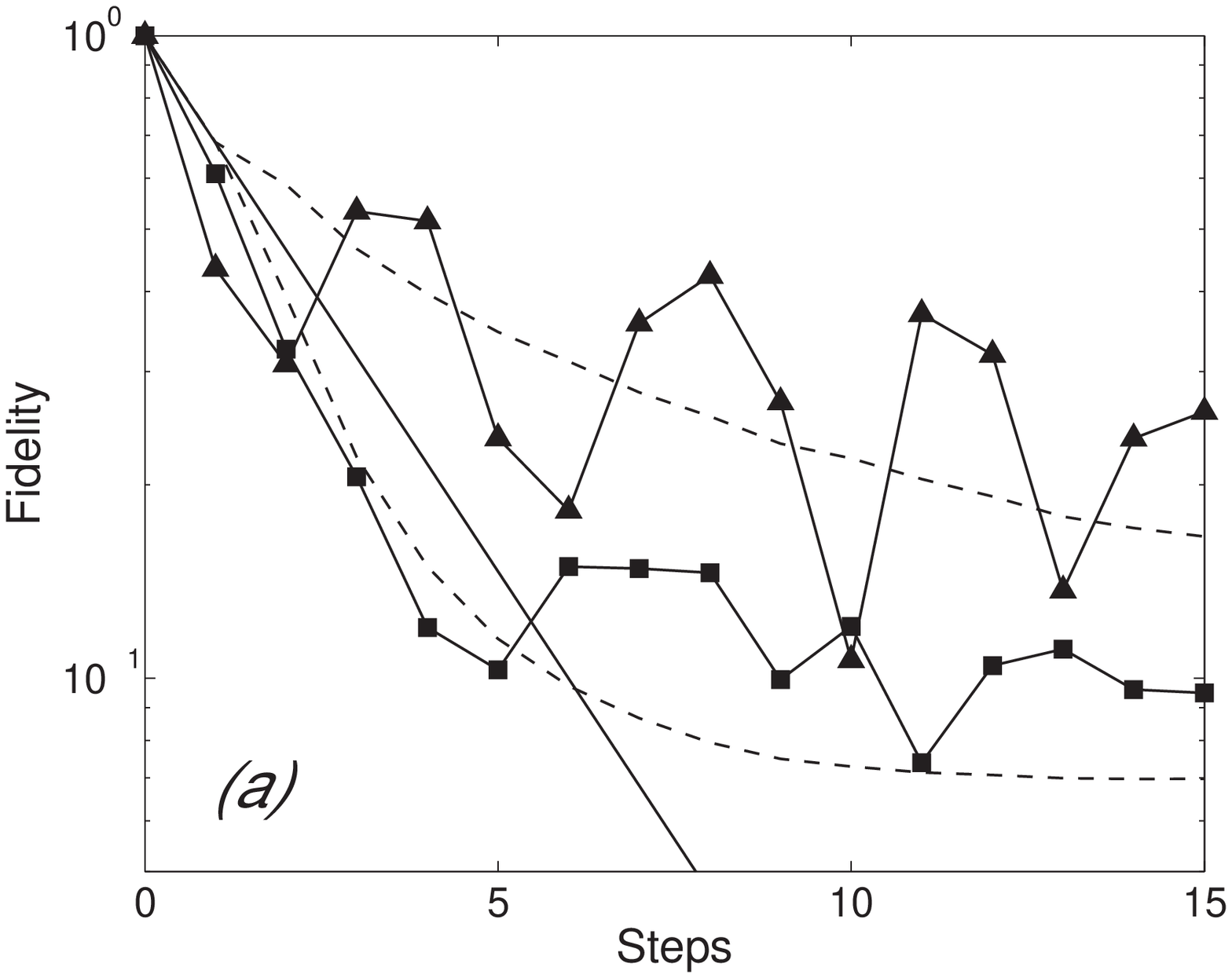}\\
\includegraphics[scale = 0.45]{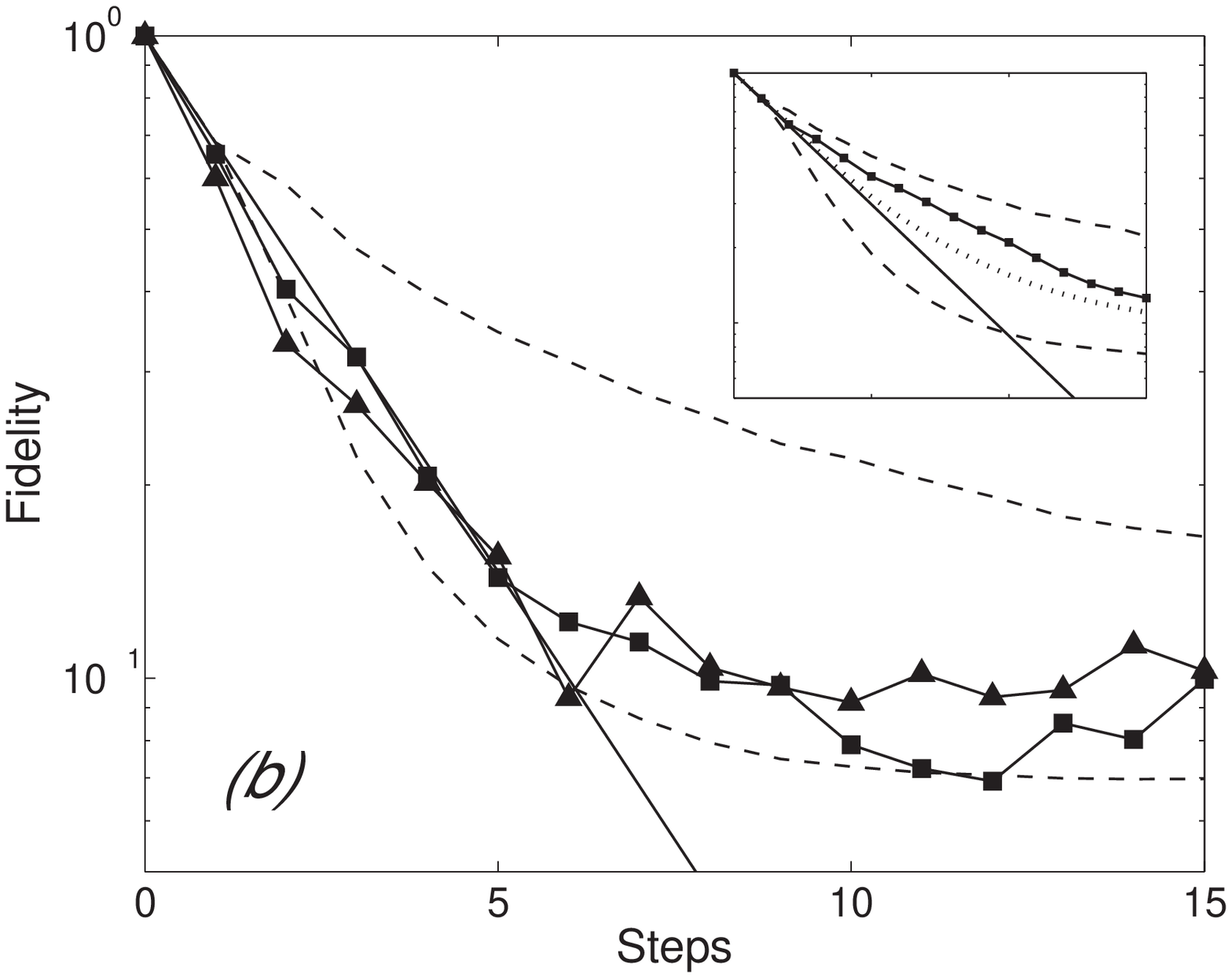} \\
\end{array}$
\caption{\label{XvsZregandrand}
Comparison of fidelity decay curves for regular (a) versus chaotic (b) evolution under different forms of perturbation.  The natural evolution of the molecule provided the regular system and a pseudo-random map was used to model chaotic evolution.  The perturbation form was controlled by the presence or absence of conjugate rotations on either side of the coupling interaction between the probe and system.  In the case of regular evolution, when the perturbation is a rotation about the z-axis ($\blacktriangle$ curves), which commutes with a system coordinate, the fidelity decay shows substantial fluctuations away from the average.  Changing the perturbation to a rotation about the x-axis ($\blacksquare$ curves) substantially alters the form of the decay.  In the chaotic case, the evolution looks random in the eigenbasis of  either perturbation and the decays exhibit the universality of the fidelity decay for complex dynamics.  The  FGR result (solid curves) and standard deviations (dashed lines) calculated from numerical simulations assuming perfect control are also shown.  The inset of figure (b) shows the experimental average fidelity decay from twenty different random maps.  The dotted line show the corresponding average from the numerical simulations and we can see that deviation from the exponential decay is also present in the numerical results and well explained by the finite size saturation level. }
\end{figure}

Different random maps can be generated by varying the pulses and delay times chosen.  This allows an averaging over random maps which is important because of the finite size fluctuations in the fidelity decay.   The average fidelity decay depends only on the relative randomness between the system and perturbation.  But, the fidelity decay for any particular map will fluctuate (an effect independent of the initial state which is already averaged over by this circuit implementation).  These fluctuations disappear as the size of the system increases beyond a few qubits, but for small systems, this effect is pronounced and requires an averaging over different random maps to obtain good statistics.  To quantify the fluctuations, numerical simulations of the fidelity decay experiment were performed over 1000 different random maps created assuming perfect control.  Averages and deviations were then calculated to compare with the experimental results.  

The results of Fig. \ref{XvsZregandrand} also show how the form of the perturbation effects the decay.  The pseudo-random map shows a universal response, in that the decays are identical under different perturbations.  On the other hand,  the natural evolution's decay varies wildly.  This result demonstrates the necessity of choosing the perturbation carefully in the context of distinguishing between regular and chaotic evolution: fidelity decay will provide useful information only if the applied perturbation commutes with the system's coordinate \cite{Emerson:2002a}.

These experiments also highlight the relevance of these techniques to decoherence studies.  For this purpose, we consider the pseudo-pure qubit to be the system we are interested in and the maximally mixed bottom register, the environment.   The decoherence rate of the system is typically governed by some {\em macroscopic} parameters of the environment, such as its temperature, cutoff frequency, etc., see e.g. Refs.~\cite{Caldeira:1985a,PZ01a}.  Recently, the importance of the {\em dynamics} of the environment has been expressed by a few authors \cite{Poulin:2004a,Kohout:2003a}, and our experiments constitute a direct demonstration of these effects because we can directly control the complexity of the environment's dynamics.  

Consider a Hamiltonian describing the evolution of the probe and environment together as $H = H_E + A_S \otimes B_E$ where $H_E$ is the self Hamiltonian of the environment and $A_S\otimes B_E$ is the coupling between the environment and the probe. We assume an initial product state $\rho_S(0) \otimes \rho_E(0)$. We obtain the state of the probe $\rho_S(t)$ by tracing out the environment after a time $t$. When expressed in the eigenbasis of the coupling ($A\ket j = \lambda_j \ket j$), the diagonal elements of $\rho_S(t)$ are equal to those of $\rho_S(0)$, while its off-diagonal terms have decayed  by a factor $\gamma_{jk}(t)$ given by (using the Trotter decomposition and $\delta \rightarrow 0$)
\begin{equation*}
\frac 1N Tr\left\{  \left(e^{i\delta (H_E + \lambda_jB)}\right)^{t/\delta} \left(e^{-i\delta(\lambda_j-\lambda_k)B}e^{-i\delta (H_E+\lambda_jB)}\right)^{t/\delta} \right\}
\label{decoheqn}
\end{equation*}        
which is identical to the trace element of Eq. \ref{traceeqn}, and so was directly measured in our experiments. Hence, the decoherence rate in this model depends on how random the environment's dynamics appear in the eigenbasis of the coupling. Moreover, the coupling strength is modulated by the eigenvalue difference $\lambda_j-\lambda_k$, so in the FGR regime, the decoherence rate of the $(j,k)$ entry of $\rho_S$ is proportional to $(\lambda_j-\lambda_k)^2$.

\noindent{\em Conclusion} ---  Using our universal control over the qubit register we were able to simulate regular and complex dynamics for a quantum system and contrast their fidelity decay.  Our scheme avoids the scalability limitations of conventional NMR QIP by relying on neither full pseudo-pure state initialization nor full tomography.  The method is therefore scalable in the absence of noise and provides a powerful tool for characterizing the dynamics of an unknown system.  Moreover, our experiment can also be seen as a direct measurement of the decoherence rate of a system coupled to an engineered environment; wherein, universal control over the environment allows for systematic study of the impact of the environment dynamics and the coupling to the environment on the decoherence rate.   As such, it is an experimental  demonstration of the importance of the environment dynamics  in the decoherence rate. 

\begin{acknowledgments}
C.R. would like to thank M. Ditty for his technical expertise with the spectrometer.  This work was funded by NSERC, CFI, CIAR, NATEQ and ARDA.
\end{acknowledgments}

\end{document}